\documentclass{article}
\usepackage{graphicx}
\usepackage{psfig}
\begin{document}

\large
\begin{center}
{\huge{\bf Accuracy of the recovered flux of extended sources obscured by bad pixels in the central EPIC FOV}}
\\
\vspace{1.0cm}
{\Large XMM-SOC-CAL-TN-0227} \\
%Version 1.0 \\
\vspace{0.5cm}
{\Large J. Nevalainen, I. Valtchahov, R. D. Saxton and S. Molendi} \\
\vspace{0.5cm}
Feb 2 2021
\end{center}

\section{Introduction}
A fraction of the XMM-Newton EPIC FOV is obscured by the dysfunctional (i.e. bad) pixels, dominated by the CCD gaps, which complicates the analysis of extended X-ray sources.
Before SAS version 17.0.0 the only standard option for attempting the recovery of the obscured flux was to scale the measured flux by the fraction of the obscured detector area.  
This is accurate only in the very limiting case of an extended source with a uniform spatial flux distribution.

Since version 17.0.0 SAS has included an option for recovering the lost fraction of the flux obtained by a primary instrument (e.g. pn) by utilising the information of the spatial distribution of the flux
(i.e. an image) of the same source observed with another instrument (e.g. MOS2).  Assuming that the supplementary image is minimally obscured at the spatial locations where the primary instrument is obscured, the correction can be accurate.

The procedure is implemented to \texttt{arfgen}  whereby the effective area associated with the input spectrum is reduced by the lost flux fraction in the intended full extraction region, as informed by the supplementary image.
While it is the flux, rather than the effective area, that is reduced in reality, the above correction works in practice. 
Namely, when using the effective area, reduced by the above procedure, in the consequent spectral fitting of the data reduced by the bad pixel obscuration by the same factor, the net effect is the increase of the spectral emission model normalisation
compared to that obtained with the original effective area. The increased model normalisation corresponds to the full unobscured extraction region, as desired.

Since the fraction of the obscured area may exceed 10\% in the EPIC FOV (see below) the knowledge of the accuracy of the recovery procedure is important if one is interested in a high level of accuracy of the recovered flux of an extended source in the full intended extraction region. Our aim in this note is to evaluate the accuracy achieved by the recommended default procedure using a sample of clusters of galaxies. We do not examine here the accuracy of the recovery of the spectral shape since its change is expected 
to be small in general compared to the flux effect.

\section{Data}
\subsection{The sample}
We used a sample of 27 XMM-Newton observations of nearby (z $<$ 0.1) clusters of galaxies for our work (see Table 1). Galaxy clusters are useful because their stable, bright and diffuse X-ray emission fills completely the central regions of the EPIC FOV, providing sufficient numbers of counts within the pixels of our adopted size of 5 arcsec (see Section \ref{image}).
Also, the surface brightness of clusters decreases quite rapidly with the distance from the cluster center (see Section \ref{modelimages}) which is essential for the testing of the recovery of the obscured non-uniform flux.
We limit the study within 6 arcmin distance from the cluster center in order to maintain the background below 10\% of the cluster emission in the adopted 0.5-7.0 keV energy band.

\subsection{Processing}
\label{processing}
When processing the data and extracting products for generating the effective area files (i.e. event files, spectra and images), we intended to mimic the choices of a general user for wider applicability of our results,
i.e. we use default procedures and parameter values.
We processed the pn and MOS event files with \texttt{epchain} and \texttt{emchain}  procedures, respectively,  in SAS18.0.0 with default parameters.
Relevant for our analysis, the \texttt{epchain} default setting \texttt{runbadpixfind=Y} will cause the \texttt{badpixfind} routine to be executed.
In the default mode 1 (enabled by \texttt{searchbadpix=Y}) hot, dead and flickering pixels in a given observation are identified.
In case of  \texttt{emchain}, the default setting (\texttt{badpixfindalgo=EM}) activates a more sensitive \texttt{embadpixfind} routine, instead of \texttt{badpixfind} routine used for pn.

The default settings \texttt{runbadpix=Y} for \texttt{epchain} and \texttt{withbadpix=yes} for \texttt{emchain} activate the \texttt{badpix} routine with \texttt{getnewbadpix=Y} whereby the bad pixels identified above  will be propagated to the BADPIX extension of the event files.
The default settings \texttt{getuplnkbadpix=Y}  and \texttt{getotherbadpix=Y} propagate the uplinked and non-uplinked bad pixels to the BADPIX extension.

We then filtered the pn and MOS event files with the recommended expressions  
\begin{verbatim}
FLAG==0 && PATTERN<=4
\end{verbatim}
and 
\begin{verbatim}
#XMMEA_EM && PATTERN<=12
\end{verbatim}
We further filtered the event files by removing the data obtained during flare periods. 
These event files (e.g. pn.fits) will be used as arguments for the \texttt{badpixlocation} parameter when running \texttt{arfgen} in Section \ref{image}.

We extracted spectra with default parameters within central r = 6 arcmin region (defined in sky coordinates) centered at the X-ray peak (see Table 1).
We did not exclude point sources except in the case of A2142, where an exceptionally bright source is partially obscured by a pn CCD gap.
These spectra (e.g. pn.pha) will be used as arguments for the \texttt{spectrumset} parameter when running \texttt{arfgen} in Section \ref{image}.

\begin{figure*}
\begin{center}
\begin{minipage}{3in}
%\hspace*{1.5in}
\hbox{
\includegraphics[height=4.5cm]{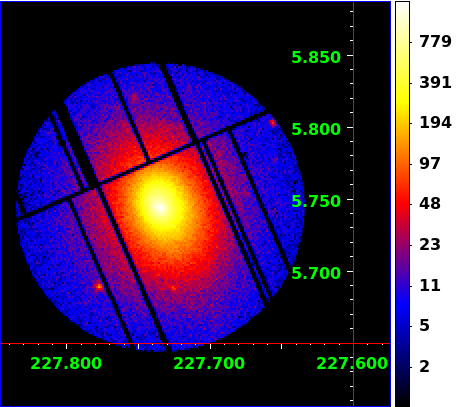}
\includegraphics[height=4.5cm]{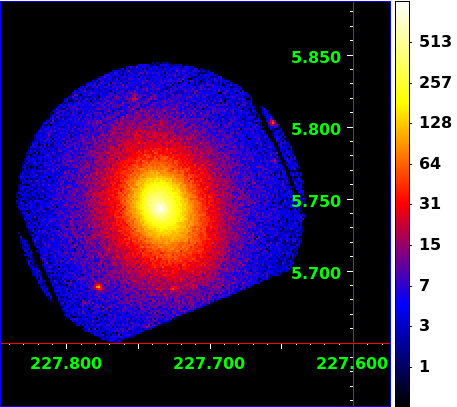}
\includegraphics[height=4.5cm]{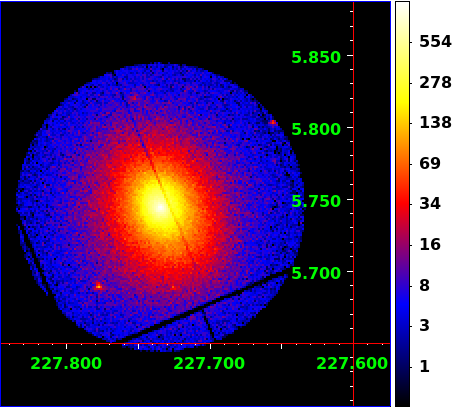}
}
\end{minipage}
\end{center}
\caption{\footnotesize
Raw count images of A2029 cluster (OBSID = 0551780301) using pn (left panel), MOS1 (middle panel) and MOS2 (right panel) in sky coordinates in the central r = 6 arcmin region. The color bars indicate the level of counts per pixel.
}
\label{raw.fig}
\end{figure*}

We extracted raw count MOS2 images within central r = 6 arcmin region in the sky coordinate system with a bin size of 5 arcsec in the 0.5-7.0 keV band.
These images (e.g. MOS2.im, see Fig. \ref{raw.fig}) will be used as arguments for the  \texttt{detmaparray} parameter when running \texttt{arfgen} in Section \ref{image}.

In order to obtain estimates of the geometric fraction R$_{area}$ of the unobscured area to that of the full intended circle with r = 6 arcmin, we utilised 
the \texttt{backscale} - routine with  \texttt{withbadpixcorr=yes}, \texttt{badpixelresolution=1} (1 arcsec) and \texttt{badpixlocation} set to the event file produced above (e.g. pn.fits).
The computations yielded that on average the pure geometric area reduction factor due to bad pixel/CCD gap obscuration (1 -$<$R$_{area}>$) is 14\% for pn, 5\% for MOS1 and 4\% for MOS2 for our sample (see Table 1). 

% Plots of A2029

% Results table
\begin{table}[t]
{ \footnotesize
\begin{center}
\caption{{\bf Results}}
\vspace{0.2cm}
\begin{tabular}{lcccccccccccccc}
\hline
cluster    & Obs. ID    &  RA        & DEC        & \multicolumn{3}{c}{R$_{area}$$^{a}$}           &  R$_{image}$$^{b}$ &  R$_{model}$$^{c}$ \\   
           &            &            &            & \multicolumn{3}{c}{------------------------}  &                  &                   \\ 
           &            &            &            &  pn   & M1     & M2                           &                  &                    \\ 
\hline 
A85        & 0723802101 & 10.460333  & -9.303008  & 0.866 & 0.932  & 0.963                        & 0.8819           & 0.8851 \\   
A85        & 0723802201 & 10.460333  & -9.303008  & 0.866 & 0.933  & 0.965                        & 0.8807           & 0.8846 \\ 
A401       & 0112260301 & 44.740833  & 13.582642  & 0.862 & 0.960  & 0.966                        & 0.8564           & 0.8496 \\ 
A478       & 0109880101 & 63.354875  & 10.465269  & 0.860 & 0.965  & 0.972                        & 0.8828           & 0.8898 \\
A644       & 0744412201 & 124.356792 & -7.512367  & 0.867 & 0.931  & 0.960                        & 0.8962           & 0.8837 \\
A754       & 0136740101 & 137.333375 & -9.679144  & 0.849 & 0.969  & 0.972                        & 0.8607           & 0.8595 \\
A1650      & 0093200101 & 194.673000 & -1.761561  & 0.847 & 0.958  & 0.958                        & 0.8327           & 0.8287 \\
A1651      & 0203020101 & 194.843583 & -4.196039  & 0.867 & 0.960  & 0.965                        & 0.8758           & 0.8811 \\
A1795      & 0097820101 & 207.220833 & 26.590278  & 0.887 & 0.962  & 0.965                        & 0.9087           & 0.9035 \\
A2029      & 0551780301 & 227.734167 & 5.744639   & 0.871 & 0.899  & 0.963                        & 0.9312           & 0.9278 \\
A2142      & 0674560201 & 239.583208 & 27.233158  & 0.865 & 0.945  & 0.951                        & 0.8659           & 0.8742 \\
A2244      & 0740900101 & 255.676958 & 34.060236  & 0.866 & 0.929  & 0.961                        & 0.8906           & 0.8917 \\
A2319      & 0302150101 & 290.297333 & 43.949839  & 0.863 & 0.960  & 0.960                        & 0.8752           & 0.8779 \\
A2319      & 0302150201 & 290.297333 & 43.949839  & 0.873 & 0.859  & 0.965                        & 0.8901           & 0.8895 \\
A2319      & 0600040101 & 290.297333 & 43.949839  & 0.866 & 0.954  & 0.962                        & 0.8698           & 0.8643 \\
A3112      & 0603050101 & 49.490417  & -44.238417 & 0.866 & 0.947  & 0.960                        & 0.9070           & 0.9090 \\
A3391      & 0505210401 & 96.585067  & -53.693201 & 0.866 & 0.959  & 0.967                        & 0.8511           & 0.8622 \\
A3558      & 0107260101 & 201.987125 & -31.495583 & 0.868 & 0.966  & 0.966                        & 0.8750           & 0.8788 \\
A3571      & 0086950201 & 206.867667 & -32.864533 & 0.867 & 0.965  & 0.968                        & 0.8745           & 0.8713 \\
A3827      & 0149670101 & 330.469708 & -59.946483 & 0.865 & 0.962  & 0.969                        & 0.8723           & 0.8695 \\
Coma       & 0300530101 & 194.954625 &  27.960983 & 0.858 & 0.906  & 0.960                        & 0.8686           & 0.8590 \\
Coma       & 0153750101 & 194.954625 &  27.960983 & 0.864 & 0.968  & 0.973                        & 0.8661           & 0.8665 \\
CygA       & 0302800101 & 299.868042 &  40.733919 & 0.867 & 0.954  & 0.962                        & 0.9073           & 0.9022 \\
PKS0745    & 0105870101 & 116.880417 & -19.294394 & 0.865 & 0.961  & 0.965                        & 0.9055           & 0.9146 \\
Ophiuchus  & 0505150101 & 258.115500 & -23.368714 & 0.865 & 0.955  & 0.961                        & 0.8671           & 0.8630 \\
Triangulum & 0093620101 & 249.583167 & -64.356086 & 0.854 & 0.964  & 0.970                        & 0.8525           & 0.8469 \\
ZwCl1215   & 0300211401 & 184.421333 &  3.655767  & 0.866 & 0.944  & 0.962                        & 0.8637           & 0.8696 \\
\hline
\\
\end{tabular}
\end{center}
$^{a}$ The ratio of the areas of the good unobscured region as reported by backscale - routine in SAS 18.0.0 to the full geometric extraction region (a circle with r = 6 arcmin).\\
$^{b}$ The ratio of the effective area related to the pn spectrum reduced by \texttt{arfgen} when using the MOS2 image for correcting for the bad pixels to that obtained when applying no 
correction for the bad pixels (Eq. \ref{eq1}).\\
$^{c}$ The expected value of the effective area re-scaling factor, obtained via modelling the cluster image (Eq. \ref{eq2}, i.e. \texttt{arfgen} is not involved). 
}
\label{res.tab}
\end{table}

\section{Testing}
Since the obscured fraction of the central r = 6 arcmin area is larger for pn than for MOS (see Table 1 and  Fig. \ref{raw.fig}), the inaccuracy of the total recovered flux is probably largest for pn.
We performed our test only for the pn flux recovery to get a single, conservative estimate for the accuracy of the procedure for all EPIC instruments.

\subsection{Re-scaling factors (R$_{image}$)}
\label{image}
The relative accuracy of the recovered total flux is given by the accuracy of the re-scaling factors (R$_{image}$). Our goal in this work is to evaluate the accuracy of R$_{image}$ estimated by \texttt{arfgen} for recovering the obscured pn flux using MOS2 images.
For this we first produced the effective area files using syntax:

\begin{verbatim}
arfgen spectrumset=pn.pha arfset=pn.arf detmaptype=dataset 
detmaparray=MOS2.im withbadpixcorr=yes badpixlocation=pn.fits 
badpixmaptype=dataset extendedsource=yes
\end{verbatim}
where pn.pha, pn.fits and MOS2.im are the pn spectrum, pn event file and MOS2 image, respectively, produced as described in Section \ref{processing}.
We applied the default bad pixel resolution of 2 arcsec.
We used the resulting effective area files (pn.arf) for each observation to obtain the re-scaled effective areas A$_{eff,image}$ as a function of the photon energy.

In order to evaluate the effective area re-scaling factors estimated and implemented by \texttt{arfgen} in the above run, we also generated the effective areas A$_{eff,ref}$ by applying no bad pixel correction
(\texttt{badpixcorr=no}) 
for each observation. We then computed the re-scaling factors
\begin{equation}
 R_{image} = A_{eff,image} / A_{eff,ref}
\label{eq1}
\end{equation}
as a function of photon energy.
In our adopted 0.5--7.0 keV band, the ratios are almost constant (the standard deviation $\sim$0.1\%).
We adopted the mean R$_{image}$ value in this band as the re-scaling factor for a given observation (see Table 1).

On average the pn flux is reduced by $\sim$12\% due to the bad pixel/CCD gap obscuration in our sample, as estimated by \texttt{arfgen}, using MOS2 image for the correction. 
This is a bit smaller than the purely geometric area obscuration fraction (14\%, see Section \ref{processing}). This is due to the radially decreasing surface brightness of galaxy clusters which renders the 
loss of the detector area at larger radii less important for the total flux.

\subsection{The ``true'' values (R$_{model}$)}
We need to compare the effective area ratios R$_{image}$ obtained above to the "true" values (R$_{model}$) so that we can test the accuracy of the procedure.
We obtained these by generating 1) a model pn image of a given cluster so that we know a priori the unobscured count rate at each pixel in the full intended extraction region and 2) the spatial mask of pn bad pixels
so that we know which pixels of the model image we should reject, in order to model the bad pixel obscuration effect.

\subsubsection{The masks}
\label{masks}
When modelling the bad pixel obscuration we must make a choice of the pixel size of our masks and model images.
In order to accurately 1) map the bad pixels and 2) match the spectrum extraction region (Section \ref{processing}) 
we adopted the smallest useful pixel size of (0.05 arcsec)$^2$, which is the size of a single pixel in the internal XMM-Newton detector focal plane coordinate system.
This oversamples the physical EPIC-pn pixel by a factor of 82$\times$82.

We utilised the \texttt{eexpmap} routine of SAS for creating the spatial masks of the bad pixels. While we are not interested in the exposure maps as such (which this tool is designed to produce),
this tool anyhow provides a practical way of carrying out our task in hand. The idea is that bad pixels are traced by zero values in the exposure maps. 
In order to avoid possible complications arising when projecting the data from the detector plane (where the data are observed) to the more convenient sky coordinate system preferred by general users,
we produced our bad pixel masks in the detector coordinate system.

We thus produced pn images in detector coordinates with a pixel size of 0.05 arcsec and used those as arguments for the \texttt{imageset} parameter when running \texttt{eexpmap}.
We used \texttt{attrebin=0.021} (arcsec) to make sure that if the attitude shifts by more than half the pixel size of 0.05 arcsec 
the translation between detector and sky coordinates is recalculated.
We turned the resulting exposure maps into bad pixel masks by setting all pixels with zero exposure time as zero, while all other pixel values were set to unity
(see Fig.  \ref{model.fig}).

\subsubsection{The model images}
\label{modelimages}
\begin{figure*}
\begin{center}
\begin{minipage}{3in}
\hbox{
\includegraphics[height=5.4cm]{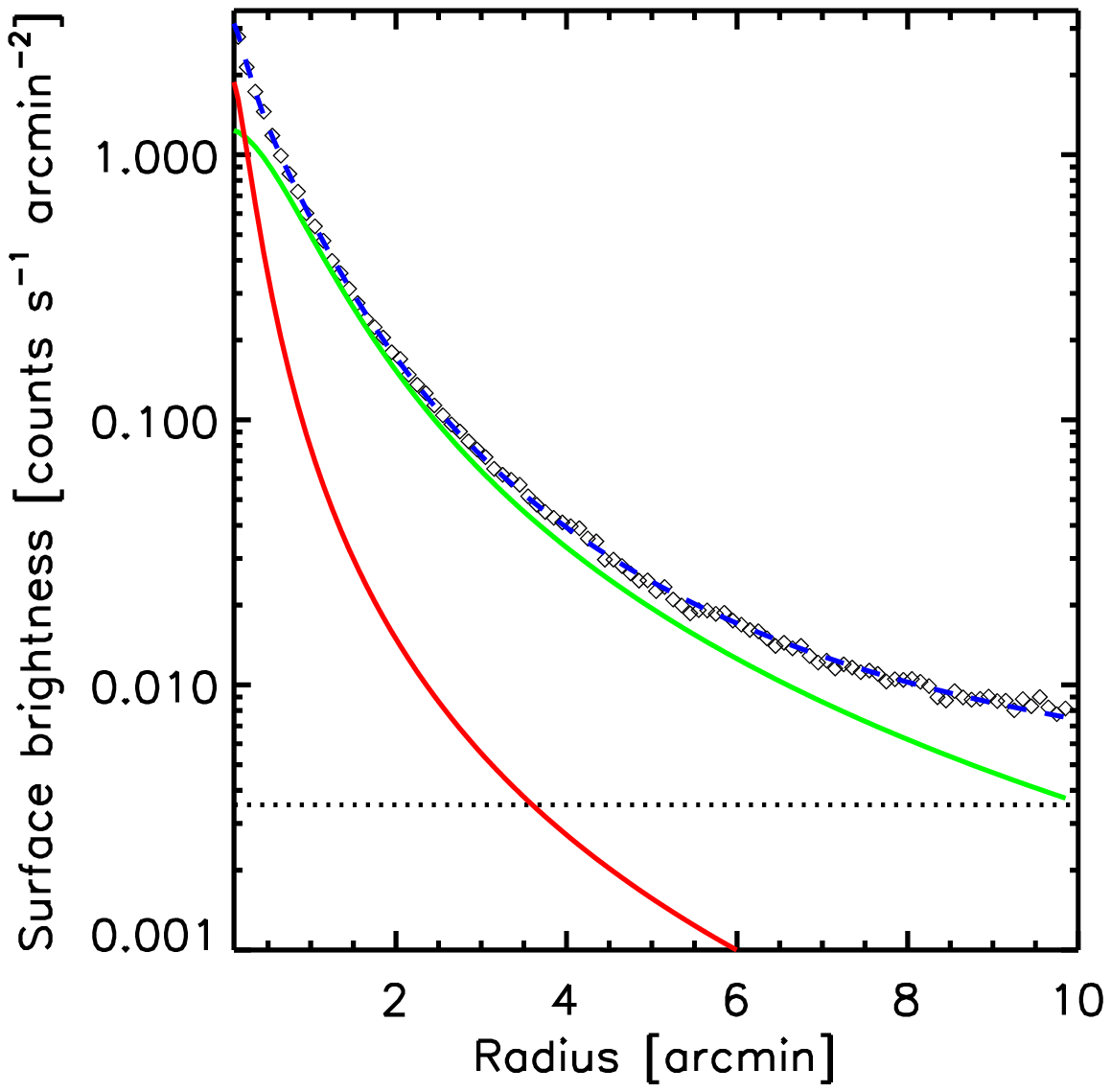}
\includegraphics[height=5cm]{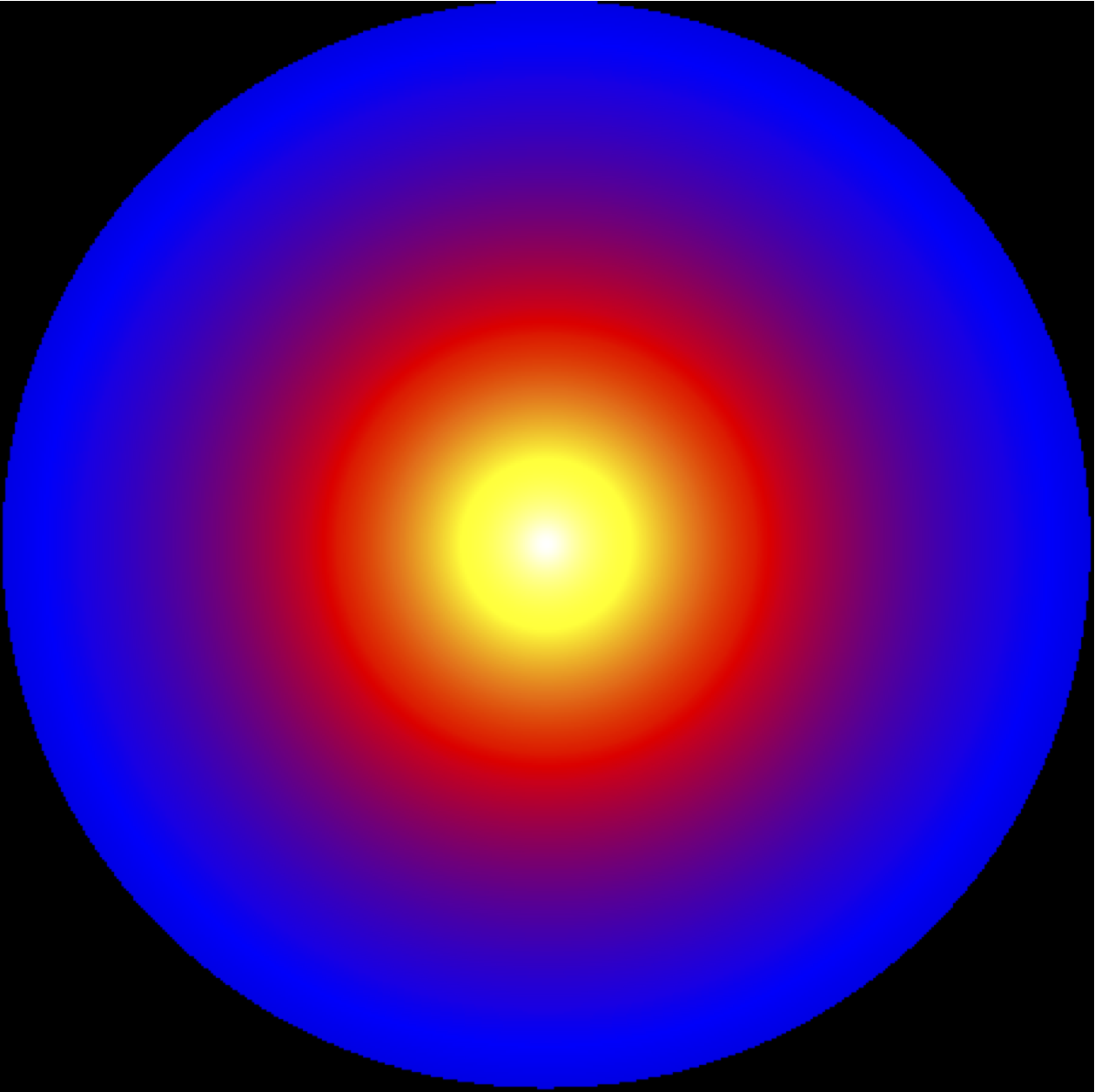}
\includegraphics[height=5cm]{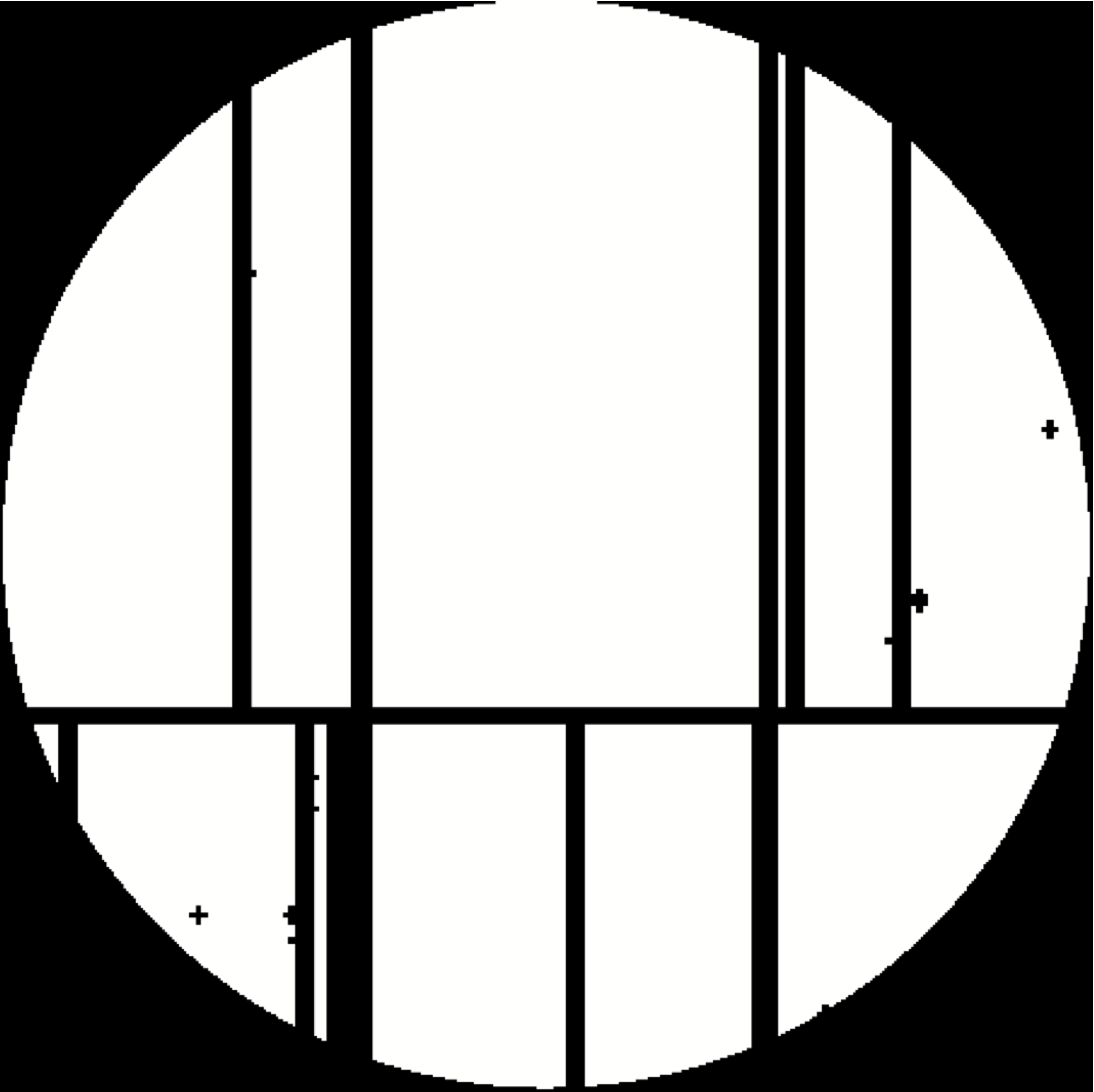}
}
\end{minipage}
\end{center}
\caption{\footnotesize
 {\it Left panel:} The MOS2 surface brightness profile data (black diamonds) of A2029 and the best-fit model (blue line) consisting of two $\beta$-model components (red and green lines) and a constant background (black dotted line).
 {\it Middle panel:} The simulated model image of A2029 using the best-fit surface brightness model within the central r = 6 arcmin region.  {\it Right panel:} the mask used for filtering the simulated image
with bad pixels (black).  
}
\label{model.fig}
\end{figure*}

We base the synthetic pn model images on the surface brightness profile models obtained by fitting the MOS2 imaging data of a given cluster.
We divided the raw count images by the exposure map to obtain vignetting-corrected count rate maps. Using those we accumulated the total count rate in co-centric annuli centered at the same point (the X-ray peak)
as the spectrum extraction region (see Section \ref{processing}) except that we ignored the bad pixels and brightest point sources. We divided the count rates with the area covered by the used pixels in a given annulus to obtain surface brightness profile. The background is defined as the constant surface brightness level at larger radii.  
We fitted the profiles with a single-$\beta$ or double-$\beta$ model
\begin{equation}
I(r) = I_{0,1} \times \left[1 + {\left( \frac{r}{r_\mathrm{core,1}} \right)}^2\right]^{(-3\beta + \frac{1}{2})} + I_{0,2} \times \left[1 + {\left( \frac{r}{r_\mathrm{core,2}} \right)}^2\right]^{(-3\beta + \frac{1}{2})} + bkg,
\label{betamodel}
\end{equation}
where r is the projected radius and bkg is the constant background (see Table 2 and Fig. \ref{model.fig})

% Plots of A2029

\begin{table}
\label{beta.tab}
  { \footnotesize
\begin{center}
\caption{{\bf Best-fit surface brightness profile model parameters}}
\vspace{0.2cm}
\begin{tabular}{lcccccccccccccc}
\hline
cluster    &  I$_{0,1}$                   &  r$_\mathrm{core,1}$ &  I$_{0,2}$                  &  r$_\mathrm{core,2}$ &  $\beta$ \\
           &  c s$^{-1}$ arcmin$^{-2}$    &  arcmin             &  c s$^{-1}$ arcmin$^{-2}$   &  arcmin             &          \\
           &                             &                     &                            &                     &          \\
\hline 
A85        &  0.32                       & 1.6                 &  1.7                       & 0.31                 & 0.53     \\
A401       &  0.09                       & 4.4                 &  0.16                      & 2.0                 & 0.89     \\
A478       &  0.88                       & 1.2                 &  2.3                       & 0.30                 & 0.65     \\
A644       &  0.11                       & 3.3                 &  0.35                      & 1.4                 & 0.81      \\
A754       &  0.02                       & 11.2                &  0.13                      & 2.4                 & 0.64      \\
A1650      &  0.13                       & 2.9                 &  0.52                      & 0.82                & 0.91      \\
A1651      &  0.14                       & 2.9                 &  0.31                      & 1.1                 & 0.90      \\
A1795      &  0.27                       & 1.7                 &  2.0                       & 0.54                & 0.59      \\
A2029      &  1.3                        & 0.95                &  2.3                       & 0.27                & 0.58      \\
A2142      &  0.43                       & 1.2                 &  0.55                      & 0.60                & 0.53      \\
A2244      &  0.40                       & 1.2                 &  0.28                      & 0.33                & 0.66       \\
A2319      &  0.15                       & 2.7                 &  0.23                      & 1.2                 & 0.49       \\
A3112      &  0.48                       & 1.1                 &  2.5                       & 0.24                & 0.65       \\
A3391      &  0.04                       & 4.4                 &  0.04                      & 1.1                 & 0.74       \\
A3558      &  0.07                       & 7.3                 &  0.19                      & 1.6                 & 1.1        \\
A3571      &  0.26                       & 3.1                 &  0.28                      & 0.77                & 0.60        \\
A3827      &  0.07                       & 4.0                 &  0.20                      & 1.6                 & 1.0         \\
Coma       &  0.14                       & 7.5                 &  ---                       & ---                 & 0.51        \\
CygA       &  3.5                        & 0.30                &  ---                       & ---                 & 0.51        \\
Ophiuchus  &  0.47                       & 4.1                 &  1.0                       & 0.46                & 0.59        \\
PKS0745    &  0.49                       & 1.2                 &  3.2                       & 0.39                & 0.68         \\
Triangulum &  0.06                       & 8.4                 &  0.23                      & 2.7                 & 0.77         \\
ZwCl1215   &  0.07                       & 3.8                 &  0.05                      & 1.5                 & 0.82         \\
\hline
\\
\end{tabular}
\end{center}
The best-fit parameters for the $\beta$ model (Eq. \ref{betamodel}) obtained by fitting the MOS2 data.
}
\end{table}

We then simulated an unobscured image of a given cluster using the surface brightness profile model obtained above (see Fig. \ref{model.fig}). As in the case of the bad pixel masks (see Section \ref{masks})
we produced the cluster images in the detector coordinate system using a pixel size of 0.05 arcsec. 
The synthetic image has identical attitude and offset between the cluster and FOV centers to those of the observational pn image and the mask image.
The synthetic image yields the model count rate $CR(model,full)$ from the unobscured full central r=6 arcmin region of a given cluster. 

The real galaxy clusters have a varying degree of deviations from the azimuthal symmetry we assumed when modelling the clusters. We essentially interpolate the surface brightness model to estimate the lost flux at the locations of the bad pixels. Given the random nature of the deviations from the symmetry we do not expect a significant bias in our modelling. This may cause some scatter between the modelled and \texttt{arfgen} - estimated fluxes which will manifest itself in our results.

\subsection{Results}
We then multiplied the model images with the masks in order to model the effect of the bad pixel/CCD gap obscuration. Using the 
mask-filtered images we then obtained the model count rate $CR(model,unobsc)$ in the surviving unobscured region. We then calculated  the ``true'' flux reduction factor, i.e. the ``true'' effective area re-scaling
factor
\begin{equation}
R_{model}  = CR(model,unobsc) / CR(model,full) 
\label{eq2}
\end{equation}
for each observation.
We then compared these with the effective area re-scaling factors R$_{image}$ (Eq. \ref{eq1}) produced with \texttt{arfgen} using the observational data (see Section \ref{image}). The comparison yielded the following results for the central r = 6 arcmin regions of the EPIC FOV using SAS 18.0.0. (see Fig. \ref{res.fig} and Table 1).

\newpage

\begin{itemize}

\item
  On average, the flux reduction fraction as given by 1) the simulations (in detector coordinate system with a pixel size of 0.05 arcsec) and by 2) \texttt{arfgen} on the observational data
  (in sky coordinates with bad pixel resolution of 2 arcsec) differ by $\approx$0.03\%

\item
  In some individual cases the difference exceeds 1\% (1.4\% at the maximum). The standard deviation of the full sample is $\approx$0.7\%.

\item
  The above results do not change significantly when improving the bad pixel resolution from the default 2 arcsec to 0.2 arcsec when running \texttt{arfgen}.
  Therefore we recommend to use the default value of 2 arcsec when analysing extended sources.

\begin{figure*}
\begin{center}
\includegraphics[height=8cm]{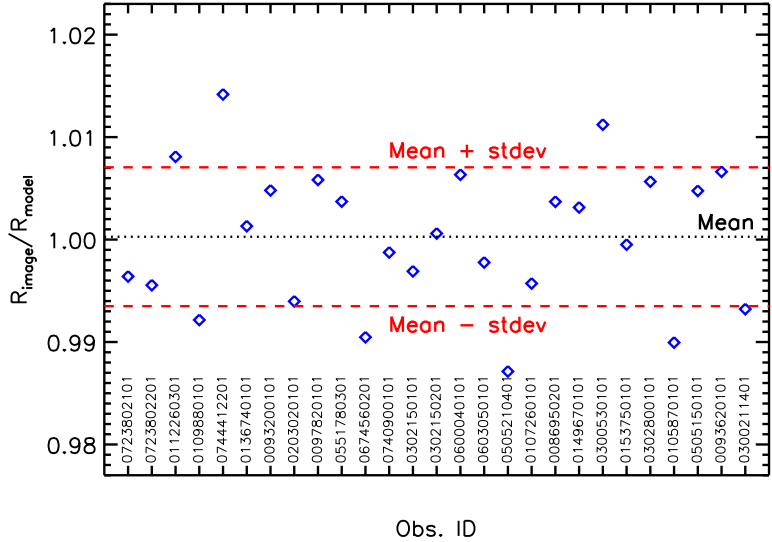}
\end{center}
\vspace*{-0.25in}
\caption{\footnotesize
  The ratios of the test values (R$_{image}$) to the model values (R$_{model}$) for each observation (blue diamonds), together with the sample average (denoted with black dotted line).
Unity corresponds to a perfect match. The red dashed lines indicate the standard deviation.  }
\label{res.fig}
\end{figure*}

\end{itemize}

\section{Conclusions}
We examined the accuracy of the recovery method of the EPIC flux obscured by the bad pixels in the central r = 6 arcmin region as implemented in SAS 18.0.0 i.e. using a supplementary raw count image
for flux information. We used a sample of galaxy clusters for testing.
We found that the accuracy of the recovered total flux is better than 0.1\% on average while in individual cases the recovered flux may be uncertain by $\sim$1\%.

{\bf Acknowledgements}\\
The work is supported by the Horizon 2020 Framework Programme of the European Union Grant Agreement No. 871158.
We acknowledge the support by the Estonian Research Council grants IUT40-2, PRG1006,  and by the European Regional Development Fund (TK133).
J.N. acknowledges ESAC Science Faculty support for a short visit when this work was initiated.

\end{document}